# REAL TIME EMULATION OF PARAMETRIC GUITAR TUBE AMPLIFIER WITH LONG SHORT TERM MEMORY NEURAL NETWORK

Thomas Schmitz and Jean-Jacques Embrechts[1]

[1]Department of Electrical Engineering and Computer Science, Liege University, Montefiore Institute, Belgium

T.Schmitz@uliege.be

jjembrechts@uliege.be

## ABSTRACT

*Numerous audio systems for musicians are expensive and bulky. Therefore, it could be advantageous to model them and to replace them by computer emulation. In guitar players' world, audio systems could have a desirable nonlinear behavior (distortion effects). It is thus difficult to find a simple model to emulate them in real time. Volterra series model and its subclass are usual ways to model nonlinear systems. Unfortunately, these systems are difficult to identify in an analytic way. In this paper we propose to take advantage of the new progress made in neural networks to emulate them in real time. We show that an accurate emulation can be reached with less than 1% of root mean square error between the signal coming from a tube amplifier and the output of the neural network. Moreover, the research has been extended to model the Gain parameter of the amplifier.*

## 1. INTRODUCTION

The modeling of nonlinear systems has been a central topic in many engineering areas, as most real-world devices exhibit nonlinear behaviors. In particular, the study of distortion effects for guitar players has been largely covered [1, 2, 3]. The reason is that musicians like the sound of tube amplifiers (in which each amplifier stage is composed of old vacuum-tube triodes). Guitarists define the sounds as more dynamic and warmer than those provided by solid state amplifiers (full transistors amplifiers). However, the tube amplifiers are often bulkier, more expensive, heavier and more fragile. This explains the large interest of the musician community for computer emulations. Even if musicians agree that these emulations get better with age, no exact correspondence between the sound coming from a tube amplifier and its emulation has been found in the literature.

In previous researches we have focused on Volterra series models [4, 5] and more specially on its subclass, the Wiener-Hammerstein cascade models [6, 7]. More specifically, researches on Hammerstein model have led to a fast Hammerstein Identification by Sine Sweep (HKISS) method [1, 2]. However, this kind of model is not sufficiently complex to correctly perform the emulation of wide range of guitar signals [1]. In this paper we propose to take advantage of the new progress made in the field of neural-networks (NN) and to evaluate the possibility of performing an accurate emulation of the *ENGL Retro Tube 50* amplifier in real time (RT).

The paper is organized as follows: the neural-network used to emulate the amplifier is presented in section 2. In section 3, the learning method and the data-set pre-processing method are described. In Section 4, the sound of the real system (i.e. the tube amplifier) and the sound from the emulated system (i.e. the NN) when a guitar signal is provided at the input are compared. Section 5 explains how to extend this model to include the amplifier's parameters (gain, equalization, ...). In this paper, the *Gain* parameter is taken as example. The effect of the knob *Gain* is to add more and more musical distortion to the guitar signal.

## 2. RECURRENT NEURAL-NETWORKS

Recurrent Neural-Networks (RNN) seem well suited to learn the nonlinear behavior of a tube amplifier. As the nonlinearities can change according to the input frequencies, it seems natural to take the previous values of the input signal *x* into account (the signal coming from the guitar) in order to compute the corresponding output signal *pred (the signal that emulate the output signal of the tube amplifier)* as depicted at Fig.1. In this case, to calculate the prediction *pred[n]* of the system, the RNN has to be fetch with a sequence of the last N values of the input signal *[x[n-(N-1)], ..., x[n]]*, where *N* is called the number of time steps (*num_step*). One can notice that the vector *h[n]* is used to compute the prediction *pred[n]*. The others *h[n-...]* vectors are used as internal states to compute *h[n]*. Their size is *num_hidden*, where *num_hidden* represents the number of hidden units in the Fully Connected layer (FC) of each cell. The main problem with RNN is its incapacity to learn the connections between two cells that are far from each other [8]. This problem is known as the *Vanishing Gradient* of deep *NN*. To avoid it, Long Short Term Memory (*LSTM*) cells have been introduced by [9]. These memory cells are used in this paper; they allow an easy propagation of long term state (see vector *c* in Fig.2) along the cells with only some minor linear interactions. The *c* vector is called the cell state; it can be interpreted as the long-term state of the cells whereas the *h* vector can be interpreted as the short-term state vector. The LSTM cell is composed of 4 FC layers. In these layers, the activation function of the neurons can be Sigmoid function σ or Hyperbolic Tangent function *tanh*. These layers interact together by gates. Considering only the *g* layer is the same as having a simple RNN cell, this layer generates a Candidate vector for the cell state. The other layers are gate controllers: *f[n]* controls which part of the cell state is kept, *i[n]* controls which part of the Candidate should be added to the long term vector *c* and finally the output gate *o[n]* controls which part of the current state should go to the output *y[n]* of this time step. Once again, one can notice that *y[n]* is not the prediction *pred[n]* of the input *x[n]*, it is the output vector at time step n. Its size is *num_hidden*. The following formatting rules must be followed strictly.

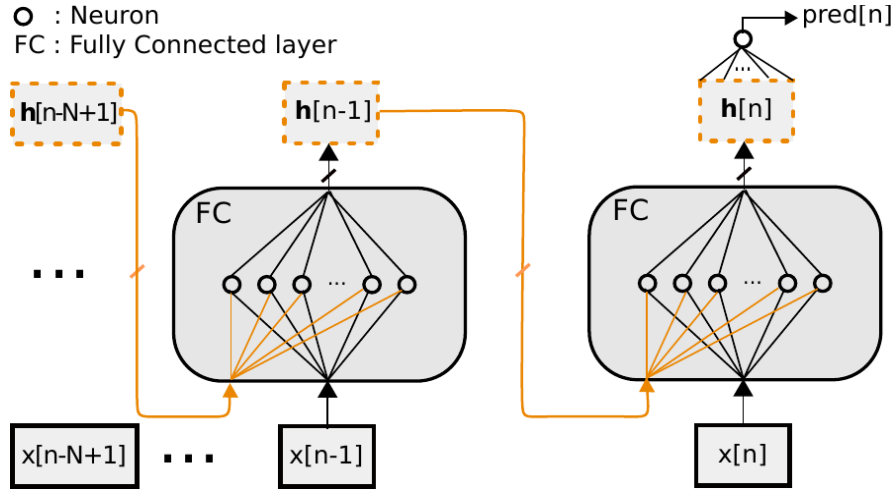

Figure 1. RNN: prediction pred[n] computed with the input sequence x of size num_step=N and the current state h[n].

### 3. LSTM APPLIED TO GUITAR SIGNAL EMULATION

The idea behind NN learning techniques is to minimize a distance (the Mean Square Error, MSE, is often taken for regression tasks) between a *target* called the *ground truth* and a *prediction*. In this case, the target is the output sample coming from the output of the amplifier that corresponds to an input sample *x[n]* coming from a guitar while the prediction is the sample *pred[n]* coming from the output of the emulator (i.e. the last LSTM cell of the layer for this same input sample *x[n])*. The learning process is based on a back-propagation algorithm [10] and gradient descent. The learning and emulating tasks can be divided in several steps: choose and format a data-set, describe a NN (called here a *graph*), execute the learning phase, save the model and use it for emulation. The description of these different phases is explained in this section. The Application Programming Interface (API) *Tensorflow* 1.3 [11] is used in this research for the description, the execution and the emulation of the graph. The source code can be found in [12].

#### 3.1. Data-set

The goal is to learn the behavior of nonlinear audio systems in order to emulate them. The choice of the signal used during the learning process is thus fundamental since it has to be representative of any guitar signal.

The input signal chosen here is a guitar signal composed of two playing techniques: some single notes and some chords (a chord is composed of several notes played at the same time). The first idea was to play each note and each chord of the guitar which resulted in a very long data-set. In fact, we experimentally found out that a data-set of twenty seconds is already long enough to bring interesting results. The data-set has to be split in 3 parts: the first one is the *Training Set*, it is used in the learning phase (gradient descent and back-propagation algorithm), the second one is the *Test Set* which will be used to evaluate the model, on data than those in the training set. Finally, the third part is the *Validation Set* which is used to check that the model has not been over-fitted by selecting convenient hyper-parameters (see Section 5). The input data which is fetched to the graph must be preliminary reshaped into 3D tensor since the LSTM input needs the following shape: *[batch_size,num_step,num_feature]*. The first dimension *batch_size* is the number of input sequences *[x[n],...,x[n-num_step-1]]* that are sent at the same time to the graph in order to compute the next gradient (this is one of the hyper-parameters). The second dimension *num_step* is the length of these sequences, it corresponds to the number of LSTM cells chained in the layer. Finally, *num_feature* is the dimension of the input signal (here, *num_feature=1* since we consider the 1D vector of audio samples of the mono signal coming from the guitar).

### 3.2. Construction of the Graph

The construction of the graph can be divided in several steps. First, the preparation of data structure (called *placeholder*) that will contain the input and the target signals. Secondly the definition of an LSTM cell as described in Fig.2. Thirdly, the cell must be unrolled over the desired number of time steps (*num_step*). Fourthly, the output vector *y[n]* has to be sent to a simple layer of neurons to reduce its dimension to a single sample prediction *pred[n]*. Fifthly, the MSE between all the predictions and the targets (*batch_size* predictions and targets) can be computed. Finally, the back-propagation and gradient descent can be applied.

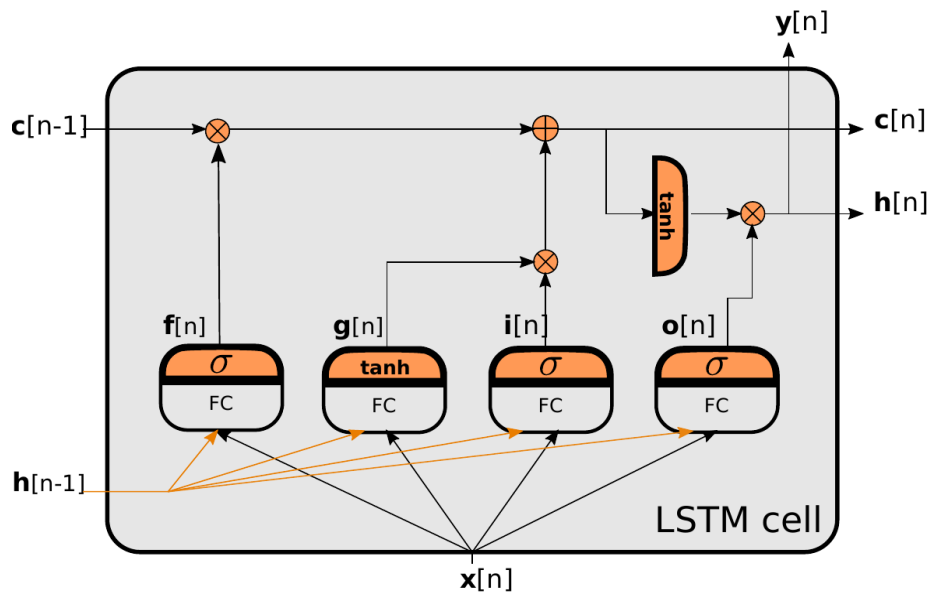

Figure 2. Long Short Term Memory cell

### 3.3. Execution of the Graph

The way Tensorflow works is first to place nodes on a graph where each node represents a mathematical operation. Then the graph is executed with special input nodes (called *placeholder*) containing input data from the data-set. The computation of the prediction starts and it is then possible to compute the MSE between the predictions and the targets. When all the data have been processed (called one *Epoch*), it restarts the computation until a satisfying level of accuracy is reached or until the accuracy do not evolve anymore. The graph and its parameters can then be saved in order to reuse it during the emulation phase. (More information is provided in the code example [12])

### 3.4. Emulation of the Graph

During the emulation phase, the graph previously saved is loaded. For real time application, the pre-processing of the guitar signal received from the sound card buffer has to be considered.

Indeed, the buffer has to be reshaped in the tensor form *[batch_size,num_step,num_feature]*. The reshaping can be efficiently carried out by the *GPU* in another graph. We can use the *batch_size* parameter as the length of the input buffer coming from the sound card. Fig.3 shows how to reshape the input data. One can notice that the *feature* parameter is equal to one, so each sample *x[n]* has to be put in a list of one element. This is due to *Python* implementation where a list *a=[$a_1$,$a_2$]* has shape=[2,] but a list *b= [[$b_1$],[$b_2$]]* has shape=[2,1]. Note also that a vector containing the last *num_step* inputs (*last buffer*) have to be stored since the values

*[x[-1] ... x[-num_step]]* are needed to compute the first values of the input tensor (see Fig.3).

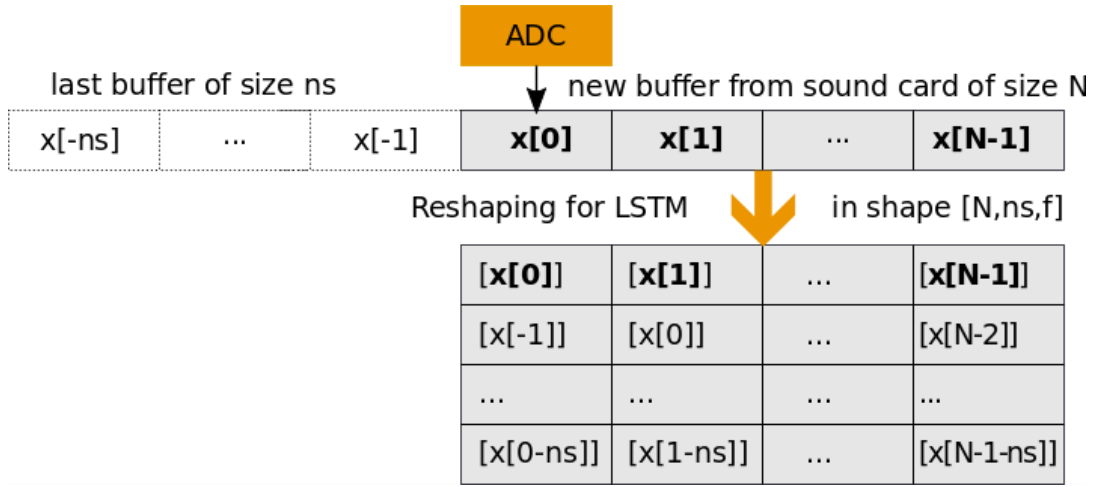

Figure 3. Reshaped input buffer of size N into LSTM input data, *ns = num_step*, f=feature=1

## 4. RESULTS

We have found that it is possible to emulate the *Engl Retro Tubes 50* at full gain (lot of distortions) with less than 1% of root mean square error RMSE between the *prediction* and the *target* (the signal that comes from the amplifier) as depicted in Fig.4 and in Fig.5 (zoom on the 800 first audio samples). The *target* signal belongs to the *validation set* and thus has never been processed by the model before. This result has been obtained with *num_step=100* and 24 hidden states. The emulation has been done by a laptop having a GPU *Nvidia gtx 1050*. As it can be seen, the curves are very close. This mean that the model is able to emulate the behaviour of the tube amplifier for a complex signal (guitar signal) that it has never seen before, in comparison with the HKISS method which can only emulate sinusoidal signals [1], this is a big improvement. The corresponding audio signals of the target and the prediction can be downloaded in *wav* format [12].

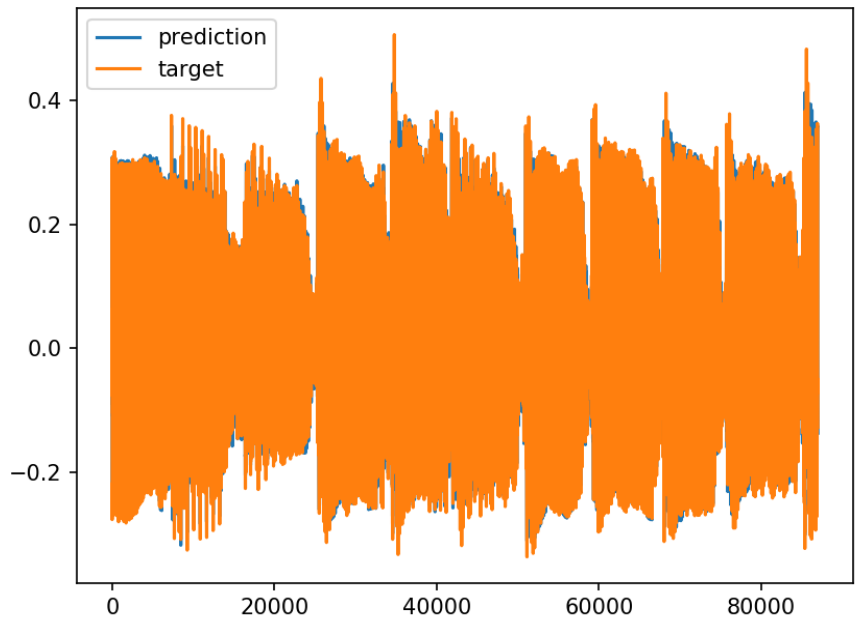

Figure 4. Temporal comparison of prediction and target signals on 2 seconds of the validation set (fs=44100Hz)

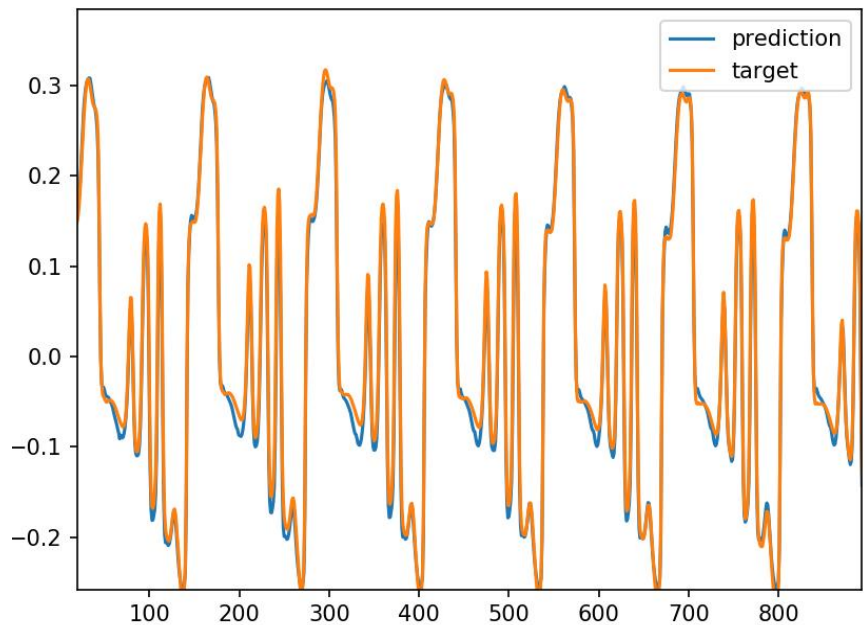

Figure 5. Temporal comparison of prediction and target signals (zoom on the first 0.02 seconds of the validation set)

## 4.1. Comparison with other models

There is no comparison to give with the HKISS method since this method does not support the emulation of such a complex signal as the guitar signal but a comparison with other NN structures can be made. With a Deep NN composed of 6 layers of 512 neurons (same input layer than in the

LSTM case) gives a RMSE of 20% which is poor. With a *Convolutional Neural Networks [13]* structure our best result was a RMSE of 16%. The LSTM model seems thus well suited for the emulation task of a tube amplifier.

## 5. MODELING OF THE PARAMETERS OF THE TUBE AMPLIFIER

In the previous section an accurate model of the amplifier *ENGL Retro Tube 50* has been build. We can go further and try to include the amplifier's parameters (usually there are at least 4 parameters, the Gain parameter which sets the amount of desired distortion and 3 equalizer's parameters: Low, Middle, Treble). An interesting property of LSTM NN is that they provide an easy way to model the effects of these parameters. Indeed, the third dimension of the LSTM 's input (*num_feature*) can be used to increase the input size of data fetched to the input of the NN. For example, a two dimensional input data (*num_feature=2*) would consist of the audio sample $x[n]$ and the gain $g[n]$ that the amplifier had during the capture of the *target[n]*. The data-set in now composed of 3 columns *[x[n],g[n],target[n]]*.

This method with the modified LSTM NN has been applied and it also gives good results with less than 1% of RMSE. However, the model is more complex and needs more hidden units and time steps to achieve this performance. This limits the possible accuracy of the model for real time applications. Several methods have been employed in order to improve the performance of the model (i.e. smaller RMSE with smaller *num_hidden* and *num_step*): batch normalization [14], Xavier and He initialization [8], dropout [15], hyperbolic tangent and *RELU* activation function [8], faster optimizer than gradient descent (AdaGrad, RMSProp, Adam) [16]. With these methods a real-time model with less than 2% of RMSE has been found with 100 time-steps and 150 hidden units.

### 5.1. Hyper-parameters Exploration

LSTM have many hyper-parameters, among them are: *batch_size, num_step, num_hidden, num_layer* which are studied by letting a well-defined function to choose them randomly and train the model during a short period (ex. 3 hours). Applying this procedure many times allows the comparison of the RMSE for different sets of hyper-parameters. To speed up the learning phase, only 3 different *Gain* parameters have been taken in our training data-set. Figs.6 and 7 give the RMSE between the *target* and the *prediction* one or two layers of LSTM cells respectively. Each figure contains 2 graphs: the first one is a 3-dimensional view of the RMSE values in the (*batch_size, num_step, num_hidden*) hyper-parameters space. The second one is a projection in a *batch_size-num_step* plane. For real-time emulation, we would be interested to minimize the number of time steps and hidden units (lower left corner of the 2D graph). Figs.6 and 7 clearly show that the RMSE decreases if the number of hidden unit increases. Concerning the time-steps, a number between 100 and 200 seems sufficient: increasing it above this value would slow down the learning without improving the RMSE. One can also notice that the model performs slightly better with two stacked layers (a layer is composed of *num_step* chained LSTM cells). Finally, it is more difficult to have a clear opinion concerning the batch size parameter. This parameter strongly depends on the GPU used to execute the model: large value of *batch_size* allows a more accurate calculation of the gradient and takes a better advantage of the parallel abilities of the GPU. A *batch_size* value around 1000 has been found for us. In conclusion, for RT applications, choosing *[num_step,num_hidden]≈[100,150]* seems fine.

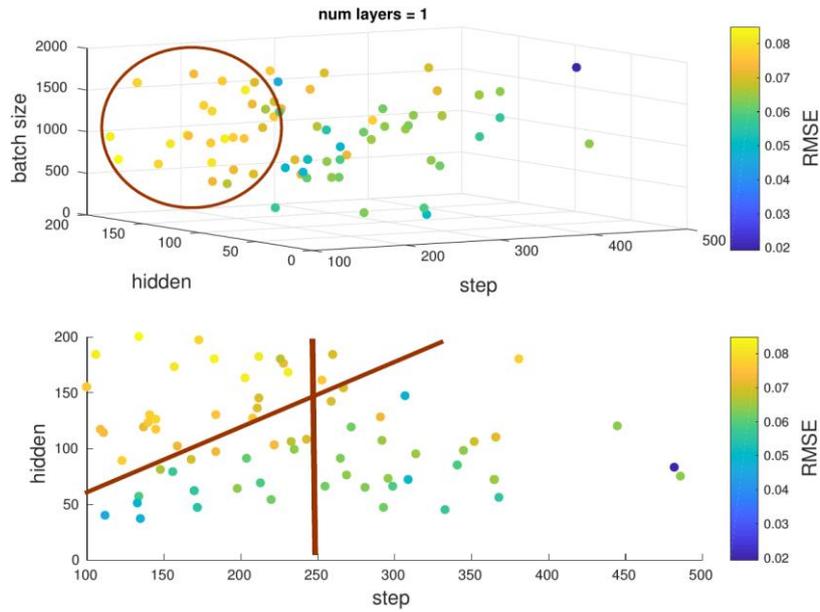

Figure 6. Comparison of RMSE between target and prediction signals using random hyper-parameters for num_layer=1

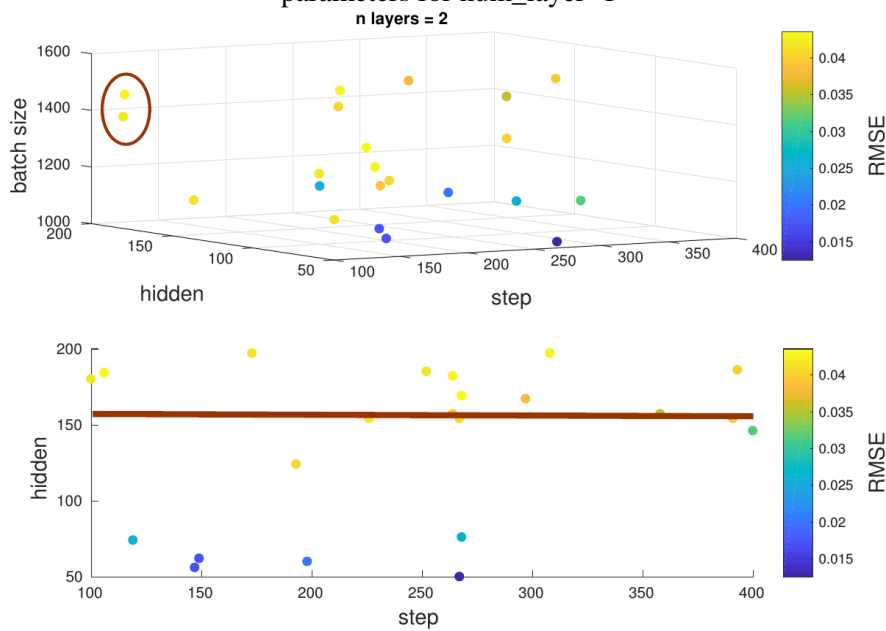

Figure 7. Comparison of RMSE between target and prediction signals using random hyper-parameters for num_layer=2.

## 6. CONCLUSIONS

LSTM and more generally NN have opened new perspectives to solve complex acoustic problems. The growing computational capability of new processors allows to run these models close to the real-time constraint which is important in many case such as for our emulation process. By its flexibility, the LSTM model has outperformed the cascade of Hammerstein model [1] which only was able to make accurate simulations of pure tone signals.

## ACKNOWLEDGEMENTS

The Titan Xp used for this research was donated by the NVIDIA Corporation.

**Authors**

Pr. J-J. Embrechts received the degree in Electrical Engineering (1981) and the Ph.D. degree (1987) from the University of Liege (ULg). Since 1999, he is a professor at the University of Liege, in the Department of Electrical Engineering and Computer Science, where he is responsible for teaching acoustics, electroacoustics, audio and video engineering and lighting techniques. He is a member of the Board of Administration of the Belgian Acoustical Society (ABAV), a member of the Audio Engineering Society (AES), the European Acoustics Association (EAA). His current research interests are in room acoustics computer models, auralization, scattering of sound waves by surfaces, microphones and loudspeakers arrays and more generally audio signal processing.

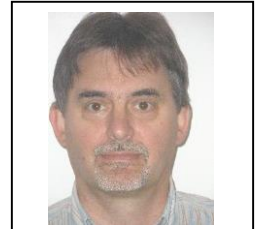

Thomas Schmitz received the degree in Electrical Engineering (2012) from the University of Liege (ULg). His final project focused on the emulation of an electrodynamics loudspeaker including its nonlinear behavior. He is presently a Ph.D. student in Laboratory for Signal and Image Exploitation (INTELSIG) research unit of the Electrical Engineering and Computer Science (EECS) department, University of Liege, Belgium. His research interests are on signal processing, nonlinear modeling, real time emulation of guitar audio systems.

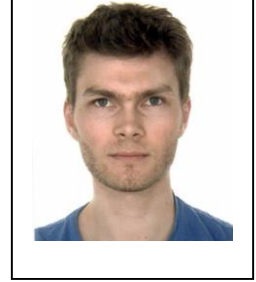